\documentclass[amsmath,amssymb,aps,pra,preprint,groupedaddress,showpacs]{revtex4-1}

\usepackage{color,graphicx}

\begin{document}

\title{Unraveling Mirror Properties in Time-Delayed Quantum Feedback 
Scenarios}

\author{Fabian M. Faulstich\textsuperscript{1,2}}
\email[E-mail me at: ]{faulstich@math.tu-berlin.de}
\homepage[Visit: ]{http://page.math.tu-berlin.de/~faulstich/}
\author{Manuel Kraft\textsuperscript{1}}
\author{Alexander Carmele\textsuperscript{1}}
\affiliation{\textsuperscript{1}Institut f\"ur Theoretische Physik, Technische Universit\"at
Berlin, Hardenbergstra\ss e 36, 10623 Berlin, Germany\\ \textsuperscript{2}Institut f\"ur Mathematik, Technische 
Universit\"at Berlin, Stra\ss e des 17.\ Juni 136, 10623 Berlin, Germany}

\begin{abstract}
We derive in the Heisenberg picture a widely used phenomenological 
coupling element to treat feedback effects in quantum optical 
platforms. 
Our derivation is based on a microscopic Hamiltonian, which describes
the mirror-emitter dynamics based on a dielectric, a mediating fully 
quantized electromagnetic field, and a single two-level system 
in front of the dielectric.
The dielectric is modeled as a a system of identical two-state atoms.
The Heisenberg equation yields a system of describing differential operator 
equations, which we solve in the Weisskopf-Wigner limit.
Due to a finite round-trip time between emitter and dielectric, we yield 
delay differential operator equations.
Our derivation motivates and justifies the typical phenomenological
assumed coupling element and allows, furthermore, 
a generalization to a variety of mirrors, such as dissipative mirrors 
or mirrors with gain dynamics.
\end{abstract}

\maketitle

\section{Introduction} 
Feedback protocols are successfully applied to stabilize
periodic processes in classical and quantum mechanical systems \cite{wiseman2009quantum,scholl2008handbook}.
In semi-classical systems Pyragas control allows
to suppress relaxation oscillations in the switch-on dynamics of
a semiconductor laser \cite{scholl2016control}.
In those systems, feedback control is modelled via a Maxwell-based
treatment of the light-matter interaction.
The paradigm for a Maxwell theory based feedback control is the 
Lang-Kobayashi model, where part of the laser output is fed back
into the laser dynamics \cite{lang1980external,bardella2016dynamic}.
Instead of self-feedback, in quantum systems measurement-based setups
of feedback control are explored.
They allow to stabilize Fock states, theoretically predicted and already
experimentally demonstrated, e.g., in cQED systems \cite{sayrin2011real}.
Quantum feedback, however, is not restricted to a read-out and 
open quantum system approach, first experiments study the 
many-photon quantum limit of feedback 
\cite{albert2011observing,hopfmann2013nonlinear,reitzenstein2003pronounced}.
These successful experimental implementations of coherent 
feedback, or non-invasive self-feedback, increase the interest
for models, which allow predictions and interpretations of the 
observed feedback effects.
A variety of models have been proposed in the linear and 
nonlinear regime.
For example, a cavity-QED system is driven into the strong coupling
regime \cite{carmele2013single}, or a laser-driven two-level system is 
partially interacting with its own emission statistics, showing modified
Mollow triplet signatures \cite{pichler2016photonic,grimsmo2015time}.
Another promising route is feedback-induced parametric squeezing
\cite{kraft2016time,nemet2016enhanced},
and enhancing of network entanglement by phase-selective feedback
based state addressing \cite{hein2015entanglement,hein2014optical}.
All these models are based on a phenomenological coupling of the 
emitters to the radiation field, namely a momentum dependent 
coupling strength.

\begin{figure}[h!]
\centering
\includegraphics{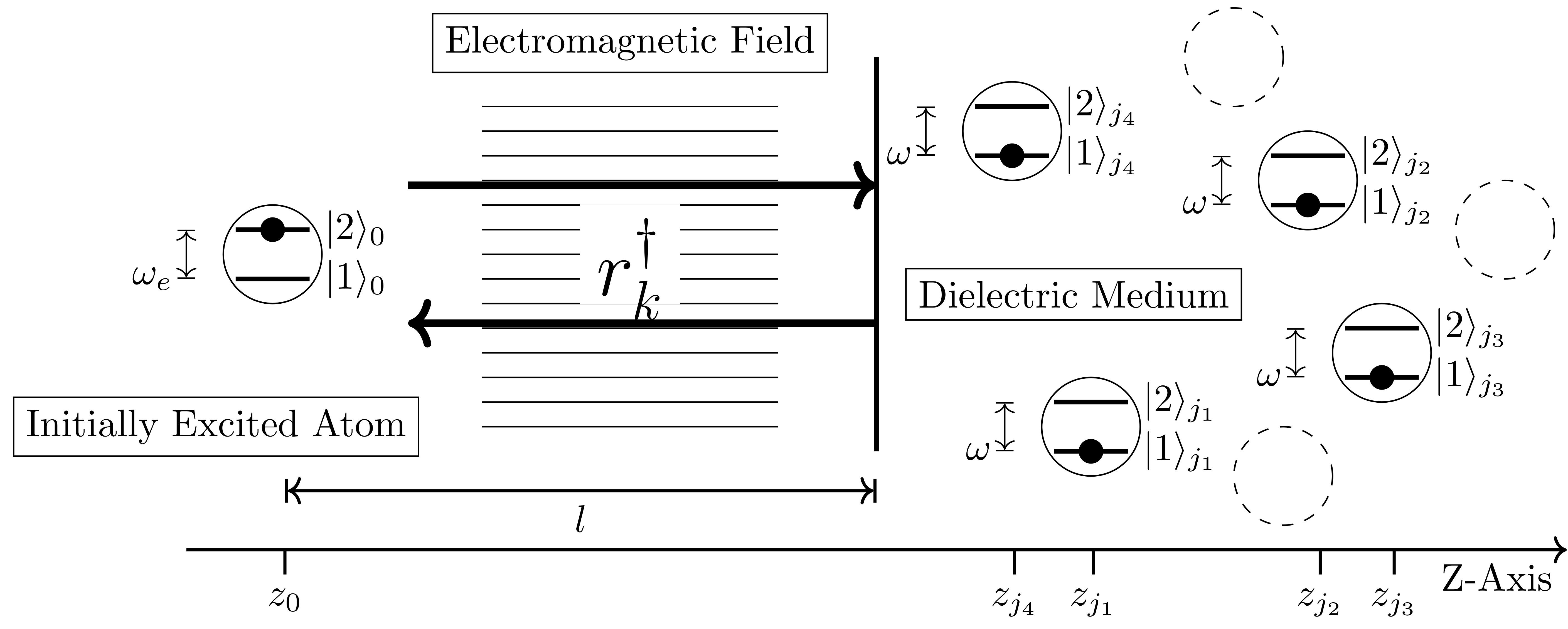}
\caption{Excited atom located in a distance $l$ to the dielectric medium.}
\label{fig:Eins}
\end{figure}

In this article, we justify the assumed coupling element.
In order to do this, we analyze the interaction of an 
initially excited two-state atom with a quantized electromagnetic 
field and a dielectric medium of $N$ two-state atoms. 
The discussion is restricted to the one-dimensional case where the electromagnetic field modes wave vectors $k$ are considered to be parallel to the $z$-axis.
This system is illustrated in Fig.~\ref{fig:Eins}.
As a system, we assume a semi-infinite one-dimensional 
waveguide \cite{hoi2015probing}. 
Material platform for such systems are, e.g., superconducting
transmission lines \cite{eichler2011experimental}, diamond nanowires
mediating between nitrogen-vacancy centers \cite{babinec2010diamond}, hollow optical fibers with cold atoms \cite{bajcsy2009efficient},
and photonic crystal waveguides coupling quantum dots \cite{laucht2012waveguide}, or plasmonic nanowires \cite{akimov2007generation}.  
The article is structured as follows. 
First, we describe briefly in the next section, Sec.~\ref{sec:HamEff}, the 
phenomenological model that is widely used in the literature \cite{kabuss2016unraveling,dorner2002laser}.
This model is our benchmark, and the following sections fulfill the 
purpose to give a microscopic justification for the applied quantum 
optical, momentum-dependent coupling.
In order to do this, we present a microscopical Hamiltonian in
Sec.~\ref{sec:MicHam}.
Starting from this Hamiltonian, we employ the Heisenberg equation 
approach and derive operator equations of motions for the dynamics 
of an atom near a plane dielectric interface, mathematically rigorously
in Sec.~\ref{sec:OpEq}.
This section ends with an effective operator equations in analogy
to the effective model of Sec. \ref{sec:HamEff}.
This is the main result of the paper.
In Sec.~\ref{sec:Concl}, we conclude the article and give a short outlook,
about possible extensions of the model.
\section{Phenomenological Model} \label{sec:HamEff}
In this section, we describe the effective and widely used phenomenological
model.
Commonly, the interaction between a dielectric and an electromagnetic field is 
simplified by assuming a hard-wall boundary at the position of the dielectric. 
This assumption yields the following Hamiltonian 
\cite{dorner2002laser,kabuss2016unraveling}:
\begin{equation}
H_{eff}/\hbar
=
\omega_e 
P^{\dag}P+\sum_k\sqrt{\frac{2cK_{0,0}}{L\pi}}\sin(kl)\left(P^{\dag}r_k+r_k^
{\dag}P\right)
+\sum_k\omega_k r_k^{\dag}r_k~,
\label{eq:HamEff}
\end{equation}
where $\omega_e$ is the transition frequency of the atom, $K_{0,0}$ is the coupling constant between atom and field, $L$ is the 
length of the quantized box,
$\omega_k$ is the frequency of mode $k$ and $l$ is the distance between the 
atom and the dielectric.
The ground and excited state of the initially excited atom at position $z_0$ 
are described by the operators $G=|0\rangle\langle0|$ and 
$E=|1\rangle\langle1|$, respectively.
Its excitation (resp. de-excitation) dynamics is denoted by the raising (resp. lowering) operator $P^{\dag}=|2\rangle\langle1|$ (resp. $P=|1\rangle\langle2|$).
The $k$-th mode of the electromagnetic field is described by the creation
(resp. annihilation) operator $r^{\dag}_k$ (resp. $r^{}_k$).
Note, the dynamics of the dielectric are here fully represented by the 
momentum-dependent coupling element $ \sin(kl) $ and do not appear
explicitly in this effective Hamiltonian.

This Hamiltonian introduces an interesting quantum feedback due to the structured continuum approach, namely by introducing a wavelength dependent coupling between the emitter and the photons of the reservoir.
Known from perturbation theory, this kind of frequency dependent coupling leads to numerous non-Markovian effects and renders Weisskopf-Wigner approaches impossible to treat quantum feedback.
Deriving the dynamics of the excitation operator via the Heisenberg equation of motion $ -i\hbar \partial_tO=[H_{eff},O] $, the feedback mechanism becomes apparent:
\begin{equation}
\begin{aligned}
\frac{d}{dt}P^{\dag}(t)
&=
-K_{0,0}P^{\dag}(t)
+K_{0,0}e^{-i\omega_e\tau}P^{\dag}(t-\tau)\Theta(t-\tau)(G(t)-E(t))\\
&\quad+\int_{\mathbb{R}}dk~i\sqrt{\frac{2cK_{0,0}L}{\pi}}\sin(kl)e^{
i(\omega_k-\omega_e)t}r_k^{\dag}(0)(G(t)-E(t))~,
\label{eq:DelayEff}
\end{aligned}
\end{equation}
where $\Theta(t)$ is the Heaviside step function and $\tau:=2l/c$ is the 
round-trip time 
\cite{kabuss2016unraveling,kabuss2015analytical,trautmann2016dissipation}.
To derive Eq.~\eqref{eq:DelayEff} a 
transformation into the rotating frame with respect to the transition
frequency $\omega_e$ was performed and without loss of generality $l\geq0$ 
was assumed. 
The first term in Eq.~\eqref{eq:DelayEff} describes the decay of the 
excited state into the reservoir proportional to the constant $ K_{0,0}$.
Additionally to this decay, after a round-trip time of $ \tau $, part of the initial signal is fed back into the dynamics of the atomic operator which is described by the second term in Eq.~\eqref{eq:DelayEff}.
Note, evaluating this equation in the many-excitation limit is a tedious
task due to the noise contributions of the third term in Eq.~\eqref{eq:DelayEff} 
and increasing reservoir-system entanglement spread. 
Possible strategies have been proposed in the Heisenberg picture 
\cite{kabuss2016unraveling}, in the quantum cascaded approach based on 
Liouvillian \cite{grimsmo2015time} or on the quantum stochastic Schr\"odinger 
equation \cite{pichler2016photonic,lu2017intensified}.\\
Subsequently, we are interested in the development of a more detailed picture
of the feedback mechanism than described by the Hamiltonian Eq.~\eqref{eq:HamEff}.
Therefore, we explicitly include the atomic dynamics of the dielectric 
induced by the field. 
The derivation is in analogy to the calculations of P. W. Milonni and R. J. 
Cook \cite{cook1987quantum}.
However, in contrast to Milonni et al., the equations are derived in the 
Heisenberg picture, to simplify the many-excitation limit.
This is rendered possible by treating the quantum noise contributions
explicitly, which is beyond the scope of the model from Milonni et al.
Our results form the backbone for further and detailed investigations by 
expanding the proposed feedback mechanism to a wider family of mirrors 
(metallic, dielectric, active, passive) and to acknowledge possible
connections to the regime of quantum optomechanics 
\cite{sudhir2017appearance,naumann2014steady}.
In particular, in the limit of a continuously distributed dielectric 
we derive the proposed $ \sin(kl) $ coupling from a reflecting medium in
distance $ l $ justifying the phenomenological ansatz.

In the following, we will derive the polarization equation of motion
in Eq.~\eqref{eq:DelayEff} with a more microscopic model, and discuss 
thereby the limits of validity for the given implementation.
\section{Microscopic Approach to Model Quantum Feedback}\label{sec:MicHam}
To derive a more general formula including the dielectric properties, 
we use a microscopic approach describing quantum feedback, applying the 
calculation of Ref.~\cite{cook1987quantum} to the Heisenberg picture. 
The interaction Hamiltonian $H_I$ for the system illustrated in Fig. \ref{fig:Eins} is in rotating wave and dipole approximation given by:
\begin{equation}
\begin{aligned}
H_I&=
i\hbar\sum_{k\in\mathbb{N}} C_{k,0}\left(P^{\dag} r_ke^{ik z_0}-Pr_k^{\dag}e^{-ik z_0}\right)\\
&\quad+i\hbar \sum_{j=1}^N\sum_{k\in\mathbb{N}} C_{k,j}\left(\sigma_{2,1}^{(j)}r_ke^{ik z_j}-\sigma_{1,2}^{(j)}r_k^{\dag}e^{-ikz_j}\right)~,
\end{aligned}
\end{equation}
where
\begin{equation}
C_{k,j}=\frac{\mu_j }{\hbar} \left(   \frac{2\pi \hbar\omega_k} {AL}  \right)^{\frac{1}{2}}
\label{eq:CouplingElement}
\end{equation}
is the frequency dependent coupling element in the light-matter interaction. 
Here, $A$ is an effective area, $L$ is the length along the $z$-axis of the 
quantized box and $\mu_j$ is the magnitude of the transition dipole moment of 
each two-state atom.
The electronic system is described via
$\sigma_{1,1}^{(j)}=|1\rangle_{j \, j} \langle1|$ 
(resp. $\sigma_{2,2}^{(j)}=|2\rangle_{j \, j} \langle2|$) 
denoting the operator of the ground state (resp. excited state)
dynamics of the $j$-th atom in the dielectric. 
Its excitation (resp. de-excitation) dynamics is described by the operator 
$\sigma_{2,1}^{(j)}=|2\rangle_{j \, j} \langle1|$ (resp. 
$\sigma_{1,2}^{(j)}=|1\rangle_{j \, j} \langle2|$). 
The transition frequency of the atom at $z_0$ is denoted $\omega_e$. 
We assume the dielectric to consist of identical atoms with resp. transition 
frequencies $\omega=\omega_1=...=\omega_N$. 
Further, we neglect contributions orthogonal to the polarization-density, 
denoted $\mathcal{P}^2_{\perp}$ \cite{richter2009two}.
As the dielectric consists of identical atoms the magnitude of the transition
dipole moment in the dielectric can be identically chosen to be $\mu_1$ which 
is not necessarily equal to $\mu_0$ the magnitude of the transition dipole 
moment of the initially excited atom in $z_0$.
The system Hamiltonian $H$ is given by the non-interacting Hamiltonian and
the interacting Hamiltonian $H_I$. Performing a unitary transformation of $H$ 
into the rotation frame with respect to $\omega_e$, the Heisenberg equation
yields the following system of differential operator equations
\begin{subequations}
\begin{align}
\frac{d}{dt}P^{\dag}
&=
\sum_{k\in\mathbb{N}}C_{k,0}e^{-ikz_0}
r_k^{\dag}
\left(G-E\right)\label{eq:SubeqHeisenberg1}\\
\frac{d}{dt}\sigma_{2,1}^{(j)}
&= i(\omega-\omega_e)\sigma_{2,1}^{(j)}
+\sum_{k\in\mathbb{N}}C_{k,1}e^{-ikz_j}r_k^{\dag}
\left(
\sigma_{1,1}^{(j)}-\sigma_{2,2}^{(j)}
\right)\quad,j\in\{1,...,N\}\label{eq:SubeqHeisenberg2}\\
\frac{d}{dt}r_k^{\dag}
&=
i(\omega_k-\omega_e)r_k^{\dag}
-C_{k,0}e^{ikz_0}P^{\dag}
-\sum_{j=1}^NC_{k,1}
 e^{ik z_j}\sigma_{2,1}^{(j)}\quad, k\in\mathbb{N}~.\label{eq:SubeqHeisenberg3}
 \end{align}
\end{subequations}
We emphasize that we have already applied the rotating wave approximation. 
Hence, we restrict the dynamics of the electromagnetic field to be 
quasi-resonant with the atomic transition frequency $\omega_e$ and, thus, 
the intensity to be sufficiently low. 
In consequence, only certain refraction and reflection coefficients are 
rendered possible, i.e., for other material and included susceptibilities
\cite{theuerholz2013influence}, the 
operator dynamics needs to be generalized to the non-rotating wave 
regime \cite{richter2009two}.
Our goal, however, is mainly to derive a $ \sin(kl) $ kind of coupling and 
for this purpose alone, we can keep this set of equations of motion as they 
already include the desired feedback mechanism.
\section{Microscopic Theory of an Atom Near a Plane Dielectric Interface in
the Heisenberg Picture} \label{sec:OpEq}
Having described the model in the previous section, we aim to deduce a valid 
feedback equation.
We start by formally eliminating $r_k^{\dag}(t)$ in Eqs.~\eqref{eq:SubeqHeisenberg1} and \eqref{eq:SubeqHeisenberg2}.
This is achieved by applying Duhamel's formula
\cite{john1982partial}
to Eq.~\eqref{eq:SubeqHeisenberg3} and substitute the obtained solution in Eqs.~\eqref{eq:SubeqHeisenberg1} and \eqref{eq:SubeqHeisenberg2}. 
For $r_k^{\dag}(t)$ we obtain:
\begin{equation}
\begin{aligned}
r_k^{\dag}(t)
&=
e^{i(\omega_k-\omega_e)t}r_k^{\dag}(0)
-\int_0^tdt'~e^{-i(\omega_k-\omega_e)(t'-t)}C_{k,0}e^{ik z_0}P^{\dag}(t')\\
&\quad-\sum_{j=1}^N
\int_0^tdt'~e^{-i(\omega_k-\omega_e)(t'-t)}C_{k,1}e^{ik 
z_j}\sigma_{2,1}^{(j)}(t')~,
\end{aligned}
\label{eq:HeisenBergBildDuhamel}
\end{equation}
which implies the differential operator equations
\begin{equation}
\begin{aligned}
\frac{d}{dt}P^{\dag}(t)
&=
-\int_0^tdt'~\sum_{k\in\mathbb{N}}C_{k,0}^2e^{-i(\omega_k-\omega_e)(t'-t)}P^{\dag}(t')\left(
G-E
\right)\\
&\quad-\sum_{j=1}^N
\int_0^tdt'~\sum_{k\in\mathbb{N}}C_{k,0}C_{k,1}e^{-ik(z_0-z_j)}e^{-i(\omega_k-\omega_e)(t'-t)}\sigma_{2,1}^{(j)}(t')\left(
G-E
\right)\\
&\quad+
\sum_{k\in\mathbb{N}}C_{k,0}e^{-ik z_0}
e^{i(\omega_k-\omega_e)t}r_k^{\dag}(0)
\left(
G-E
\right)\\
\end{aligned}
\end{equation}
and
\begin{equation}
\begin{aligned}
\frac{d}{dt}\sigma_{2,1}^{(j)}(t)
&= 
i(\omega-\omega_e)\sigma_{2,1}^{(j)}(t)
+\sum_{k\in\mathbb{N}}C_{k,1}e^{-ik z_j}
e^{i(\omega_k-\omega_e)t}r_k^{\dag}(0)
\left(
\sigma_{1,1}^{(j)}(t)-\sigma_{2,2}^{(j)}(t)
\right)\\
& -\int_0^tdt'~\sum_{k\in\mathbb{N}}C_{k,0}C_{k,1}e^{ik z_0-ik z_j}e^{-i(\omega_k-\omega_e)(t'-t)}P^{\dag}(t')\left(
\sigma_{1,1}^{(j)}(t)-\sigma_{2,2}^{(j)}(t)
\right)\\
&-\sum_{J=1}^N
\int_0^tdt'~\sum_{k\in\mathbb{N}}C_{k,1}^2e^{ik( z_J- z_j)}e^{-i(\omega_k-\omega_e)(t'-t)}\sigma_{2,1}^{(J)}(t')\left(
\sigma_{1,1}^{(j)}(t)-\sigma_{2,2}^{(j)}(t)
\right)
\end{aligned}
\end{equation}
for $j\in\{1,...,N\}$. 
We emphasize that up to this point, no further approximation have been made.
To solve this system, we use two approximations. 
We start with the narrow-band approximation, which states that the emission 
spectrum is centered around the atomic transition frequency $\omega_e$. 
Consequently, we can restrict the following analysis on a frequency interval
$[\omega_e-\upsilon,\omega_e+\upsilon]$ on which the variation of the 
coupling constants is chosen to be small. 
Therefore, the dependency of $C_{k,j}$ on the frequency $\omega_k$ is 
negligible. 
We yield the vacuum field amplitude with:
\begin{equation}
C_{k,j}\approx \frac{\mu_j }{\hbar} \left(   \frac{2\pi \hbar\omega_e} {AL}  \right)^{\frac{1}{2}}=:C_{0,j}~,
\end{equation}
for any $j\in\{0,1\}$. 
This approximation is within the range of the previously used rotating wave
approximation and therefore does not contradict previous assumptions to the 
system.
Further, we pass to the Weisskopf-Wigner approximation. 
Here, two assumptions are made. 
First, the modes of the field are closely spaced in frequency. 
Hence, we will integrate over the frequencies instead of summing.
Second, the expectation value of the integrand oscillates rapidly for very
small times $t'\ll t$. 
Therefore, there is no significant contribution to the value of the integral. 
We derive:
\begin{equation}
\begin{aligned}
\int_0^tdt'~\sum_{k\in\mathbb{N}}C_{0,0}^2 e^{-i(\omega_k-\omega_e)(t'-t)}f_1(t')
\longrightarrow
\frac{2\pi\mu_0^2\omega_e} {A\hbar c}\langle\delta\circ g,f_1\rangle_{[0,t]}
\end{aligned}
\end{equation}
and
\begin{equation}
\begin{aligned}
&\int_0^tdt'~\sum_{k\in\mathbb{N}}C_{0,j}C_{0,l} e^{-ik Z}e^{-i(\omega_k-\omega_e)(t'-t)}f_2(t')
\\
&\longrightarrow
\frac{\pi\mu_j\mu_l\omega_e} {A\hbar c}\left(
\langle \text{exp}({i\omega_e g})\delta\circ h_1,f_2\rangle_{[0,t]}+\langle \text{exp}({i\omega_e g})\delta\circ h_2,f_2\rangle_{[0,t]}
\right)~.
\end{aligned}
\end{equation}
In the above limits $g(t'):=t-t'$, $h_1(t'):=t+Z/c-t'$, $h_2(t'):=t-Z/c-t' $
and $f_{1,2}$ are arbitrary but sufficiently smooth functions. 
Here, we have introduced the constant $Z\in\mathbb{R}$ which later will be 
replaced by differences of the particle  positions along the $Z$-axis, i.e.,
$z_i-z_j$ for $i.j\in\{1,...,N\}$. 
We further used the standard notation where $\langle\cdot , \cdot
\rangle_{[0,t]}$ denotes the $L^2$-scalar product restricted on the domain 
$[0,t]$. 
For the sake of simplicity, we used calculus notation in the dual pairing,
e.g., $\text{exp}({i\omega_e g})\delta\circ h_1$ describes the composition of 
$\delta$ with $h_1$ multiplied by $\text{exp}({i\omega_e g})$.

The above notation is used to emphasize that the delta-distribution is a 
functional generated by the Dirac measure. 
Hence, the evaluation in the point $t'=0$ is only possible for a domain in which zero is an inner point. 
As this is not the case for the given domain the dual pairing of the delta
distribution and the respective functions are not well-defined.
Expanding the respective functions by the Heaviside step function yields that
the result of the dual pairing can be multiplied by any constant 
$\alpha\in[0,1]$ and is therefore not unique. 
However, the only choice ensuring the commutator relation to hold for any $t$
is $\alpha=1/2$ which 
therefore will be used subsequently. 
For further discussion we refer the reader to Appendix \ref{App: delta}.
Setting $f_1(t')=P^{\dag}(t')(G(t)-E(t))$ and 
$f_2(t')=\sigma_{2,1}^{(j)}(t')(G(t)-E(t))$ (respectively 
$f_2(t')=\sigma_{2,1}^{(j)}(t')(\sigma_{1,1}^{(j)}(t)-\sigma_{2,2}^{(j)}(t))$ 
and $f_2(t')=P^{\dag}(t')(\sigma_{1,1}^{(j)}(t)-\sigma_{2,2}^{(j)}(t))$) for 
$j\in\{1,...,N\}$, we deduce the following system of delay differential 
operator equations:
\begin{subequations}
\begin{align}
\begin{split}
\frac{d}{dt}P^{\dag}(t)
&=
-K_{0,0}P^{\dag}(t)\\
&\quad-\sum_{j=1}^NK_{0,1}e^{-ik_0l_j}\sigma_{2,1}^{(j)}(t-l_j/c)\left(
G(t)-E(t)
\right)\Theta(t-l_j/c)\\
&\quad+
K_{0,0}\Delta B ^{\dag}(0,0,t)\left(
G(t)-E(t)
\right)\end{split}\label{eq:HeisenbergbildOperatorDGLAtomX}\\
\begin{split}
\frac{d}{dt}\sigma_{2,1}^{(j)}(t)
&=-\left(i(\omega_e-\omega)+K_{1,1}\right)\sigma_{2,1}^{(j)}(t)\\
&\quad-
K_{0,1}e^{-ik_0l_{j}}P^{\dag}(t-l_{j}/c)
\left(
\sigma_{1,1}^{(j)}(t)-\sigma_{2,2}^{(j)}(t)
\right)\Theta(t-l_{j}/c)\\
&\quad-
\sum_{J\in\{1,...,N\}\setminus\{j\}}K_{1,1}e^{-ik_0l_{j,J}}\sigma_{2,1}^{(J)}(t-l_{j,J}/c)\left(
\sigma_{1,1}^{(j)}(t)-\sigma_{2,2}^{(j)}(t)
\right)~\Theta(t-l_{j,J}/c)\\
&\quad+
K_{1,1} \Delta B^{\dag}(1,j,t)\left(
\sigma_{1,1}^{(j)}(t)-\sigma_{2,2}^{(j)}(t)
\right)~,
\end{split}\label{eq:HeisenbergBildOperatorsigmaJ2}
\end{align}
\label{eq:SystemVirtualPhotonExch}
\end{subequations}
where $k_0=\omega_e /c$, $l_{j,J}:=|z_j-z_J|$, with the special case $l_{0,j}=:l_j$, ${K_{i,j}:=\pi\mu_i\mu_j\omega_e /(\hbar Ac)}$ and
\begin{equation}
\Delta B^{\dag} (l,j,t)
:=\left(\frac{LA\hbar}{2\pi^3\omega_e\mu_l^2}\right)^{\frac{1}{2}}\int_{-\infty}^{\infty} d\omega'~ e^{i(\omega'z_j/c-(\omega_e-\omega')t)}
r_{k'}^{\dag}(0)~.
\end{equation}
Eqs.~\eqref{eq:SystemVirtualPhotonExch} explicitly expose the delay effect 
in the atom-atom coupling via a mediating electromagnetic field.
Subsequently, we use the standard notation $\tau_j:=2l_j/c$ and $\tau := 2l/c$.

\begin{figure}[h!]
\centering
\includegraphics{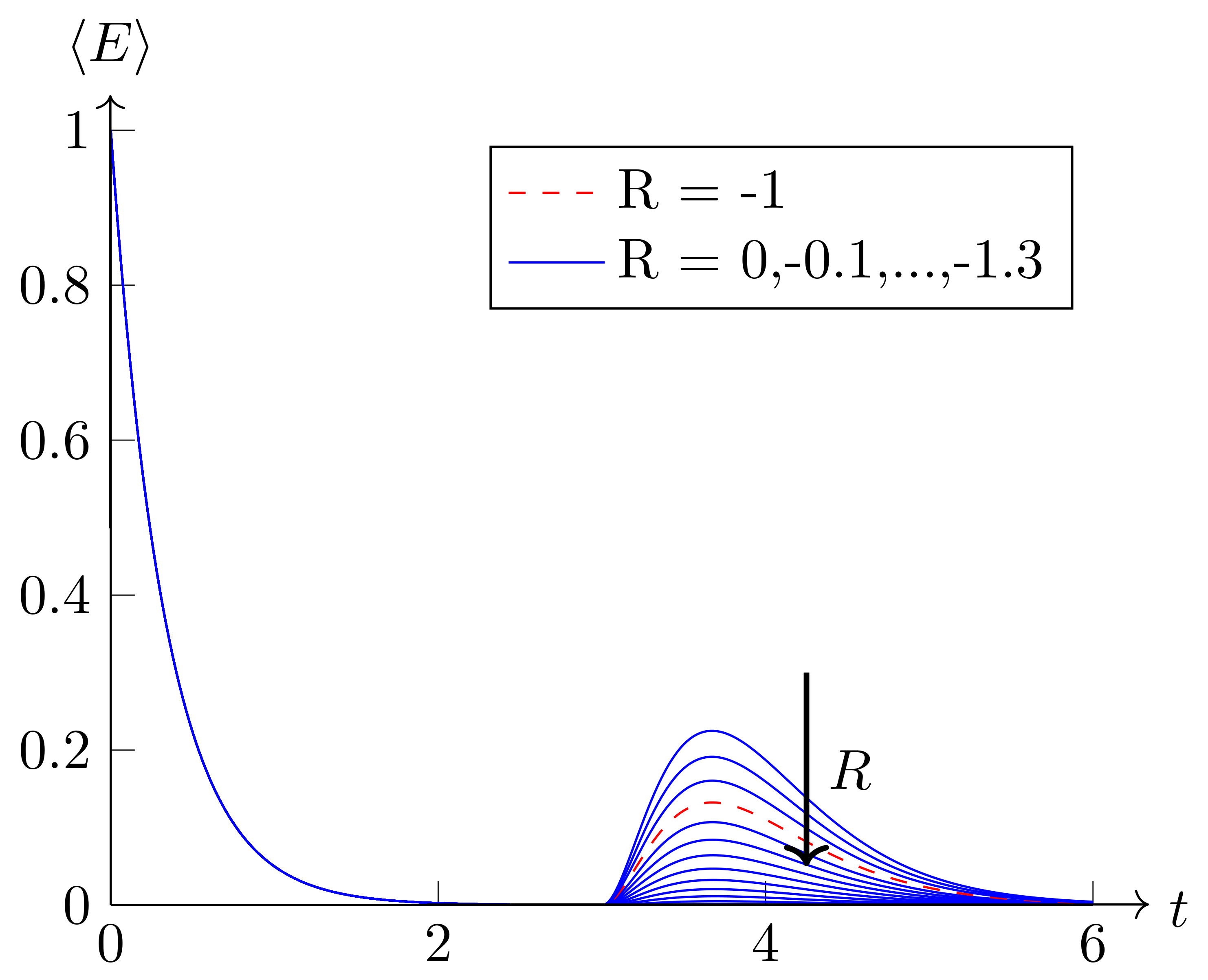}
\caption{\label{Fig:2}Expectation value $\langle E \rangle=\langle P^{\dag}P \rangle$ of the atom outside the dielectric at $z_0$ for different $R$.
The values of $R$ increase in the direction indicated by the arrow.
$K_{0,0}=1.5$, $\tau = 3$ and $\omega_e=\pi/\tau$.}
\label{fig:Zwei}
\end{figure}

Further, we consider a dielectric in which the scattered field is small
compared to the incident field on the scatterer. 
This level of treatment is consistent with the Born approximation, which 
neglects the interaction of the atoms within the dielectric via photon 
exchange.
Assuming that the dielectric and the initially excited atom 
are off-resonant yields that all atoms in the dielectric remain in their
ground state.
This and $K_{1,1}\ll |\omega-\omega_e|$ implies
\begin{equation}
\begin{aligned}
\frac{d}{dt}\sigma_{2,1}^{(j)}(t)
&=-i(\omega_e-\omega)\sigma_{2,1}^{(j)}(t)-
K_{0,1}e^{-i\omega_e\tau_{j}/2}P^{\dag}(t-\tau_j/2)
\Theta(t-\tau_j/2)+
K_{1,1} \Delta B^{\dag}(j,t)
\end{aligned}
\end{equation}
where Duhamel's formula yields the solution
\begin{equation}
\begin{aligned}
\sigma_{2,1}^{(j)}(t)
&=
e^{-i(\omega_e-\omega)t}\sigma_{2,1}^{(j)}(0)-
K_{0,1}e^{-i\omega_e\tau_{j}/2}\int_0^tdt'e^{i(\omega_e-\omega)(t'-t)}
P^{\dag}(t'-\tau_j/2)~\Theta(t'-\tau_j/2)\\
&\quad+\int_0^tdt'e^{i(\omega_e-\omega)(t'-t)}
K_{1,1} \Delta B^{\dag}(j,t')~.
\end{aligned}
\end{equation}
The following adiabatic approximation is obtained by partial integration and
using that the expectation value of the integrand oscillates rapidly, 
as assumed in the Weisskopf-Wigner approximation. 
We find:
\begin{equation}
\begin{aligned}
\sigma_{2,1}^{(j)}(t)
&=
e^{-i(\omega_e-\omega)t}\sigma_{2,1}^{(j)}(0)+
i\frac{K_{0,1}}{(\omega_e-\omega)}
e^{-i\omega_e\tau_{j}/2}P^{\dag}(t-\tau_j/2)\Theta(t-\tau_j/2)\\
&\quad-i\frac{K_{0,1}}{(\omega_e-\omega)}
e^{-i\omega_e\tau_{j}/2}e^{-i(\omega_e-\omega)(t-\tau_j/2)}P^{\dag}(0)\Theta(t-\tau_j/2)\\
&\quad+\int_0^tdt'e^{-i(\omega_e-\omega)(t'-t)}
K_{1,1} \Delta B^{\dag}(j,t')~.
\end{aligned}
\end{equation}
Eliminating the dependency of $\sigma_{2,1}^{(j)}$ in 
Eq.~\eqref{eq:HeisenbergbildOperatorDGLAtomX} by substituting the 
above solution we obtain:
\begin{subequations}
\begin{align}
&\frac{d}{dt}P^{\dag}(t) =\notag \\ 
&\quad-K_{0,0}P^{\dag}(t)-
i\frac{K_{0,1}^2}{(\omega_e-\omega)}\sum_{j=1}^Ne^{-i\omega_e\tau_{j}}
P^{\dag}(t-\tau_j)
\left(
G(t)-E(t)
\right)\Theta(t-\tau_j)\label{eq:ErsteZeil}\\
&\quad-
K_{0,1}\sum_{j=1}^Ne^{-i\omega_e\tau_{j}/2}e^{-i(\omega_e-\omega)(t-\tau_j/2)}\sigma_{2,1}^{(j)}(0)\left(
G(t)-E(t)
\right)\Theta(t-\tau_j/2)\label{eq:ZweiteZeil}\\
&\quad+
i\frac{K_{0,1}^2}{(\omega_e-\omega)}\sum_{j=1}^Ne^{-i\omega_e\tau_{j}}
e^{-i(\omega_e-\omega)(t-\tau_j)}P^{\dag}(0)
\left(
G(t)-E(t)
\right)\Theta(t-\tau_j)\label{eq:DritteZeil}\\
&\quad-
K_{0,1}K_{1,1}\sum_{j=1}^Ne^{-i\omega_e\tau_{j}/2}\notag\\
&\quad\quad\times\int_0^{t-\tau_j/2}dt'e^{-i(\omega_e-\omega)(t'-t+\tau_j/2)}
\Delta B^{\dag}(j,t')\left(
G(t)-E(t)
\right)\Theta(t-\tau_j/2)\label{eq:VierteZeil}\\
&\quad+
K_{0,0}\Delta B^{\dag} (1,0,t)\left(
G(t)-E(t)\label{eq:ZFuenftZeil}
\right)~.
\end{align}
\label{eq:GLEICHUNG}
\end{subequations}
This delay differential operator equation describes the effect of the dielectric on the polarization of the atom outside the dielectric. 
It is the central equation used to describe occupation expectation
values of the initially excited atom, i.e., the system of interest. 
We emphasize that the influence of quantum noise terms like Eq.~ \eqref{eq:VierteZeil} makes this equation difficult to handle.\\
We now deduce a formula that describes an idealized mirror system 
similar to Eq.~\eqref{eq:DelayEff}.
Analogously to the calculations in \cite{cook1987quantum},
we pass to the limit of a continuously distributed dielectric. 
This will simplify the system by eliminating the sums as we are using the
Weisskopf-Wigner approximation. 
Assuming that the dielectric contains $NA~dz$ identical atoms in the slice $[z,z+dz]$ yields
\begin{equation}
\sum_{j=1}^Ne^{-i\omega_e\tau_j}P^{\dag}(t-\tau_j)\Theta(t-\tau_j)
\rightarrow
-i\frac{NA}{2k_0}P^{\dag}(t-\tau)\Theta(t-\tau)e^{-i\omega_e\tau}~,
\end{equation}
where a coordinate system was chosen such that $z_0=0$. 
We now restrict the system Eq.~\eqref{eq:GLEICHUNG} by only taking terms that scale linearly in $K$ into account. 
Hence, we neglect \eqref{eq:ZweiteZeil}, \eqref{eq:DritteZeil} and \eqref{eq:VierteZeil}. 
The remaining differential equation is given by
\begin{equation}
\begin{aligned}
\frac{d}{dt}P^{\dag}(t)
&=
-K_{0,0}P^{\dag}(t)+
K_{0,0}\Delta B^{\dag} (1,0,t)\left(
G(t)-E(t)
\right)\\
&\quad+\frac{K_{0,1}^2}{(\omega-\omega_e)}
\frac{NA}{2k_0}e^{-i\omega_e\tau}P^{\dag}(t-\tau)\Theta(t-\tau)
\left(
G(t)-E(t)
\right)~.\\
\end{aligned}
\label{eq:FinalEQ}
\end{equation}
This delay differential operator equation is indeed similar to Eq.~\eqref{eq:DelayEff}. 
To give a more explicit similarity, we identify the Fresnel 
reflection coefficient in the rotating wave approximation 
(see Appendix \ref{App: A}) in Eq.~\eqref{eq:FinalEQ}.
Since $K_{0,1}^2=K_{1,1}K_{0,0}$, we observe
\begin{equation}
\begin{aligned}
\frac{K_{0,1}^2}{(\omega-\omega_e)}\frac{NA}{2k_0}
&=K_{0,0}
\frac{N\pi\mu_1^2}{2\hbar(\omega-\omega_e)}~.
\end{aligned}
\end{equation}
The refraction index of a dielectric of $N$ two-state atoms per unit volume, each of transition frequency $\omega$ and transition dipole moment $\mu_1$ without local field effects, in the rotating wave approximation can be characterized by 
\begin{equation}
n^2(\omega_e)-1\approx\text{Re}(\chi(\omega_e))
\approx
\frac{N\pi\mu^2_1}{\hbar(\omega-\omega_e)}~.
\end{equation}
For a more detailed derivation see Appendix \ref{App: A}. 
We emphasize that for the deduction of primary Eq.~\eqref{eq:DelayEff} a 
negligible absorption of the dielectric is assumed. 
Therefore $\chi(\omega_e)=n^2(\omega_e)-1$ is taken as real valued.
The reflection coefficient according to the Fresnel formula for normal
incidence is given by $R=-(n-1)/(n+1)$. 
In the case ${\omega \approx \omega_e}$ and ${n(\omega_e) \approx 1}$ we obtain:
\begin{equation}
R=-\frac{N\pi\mu^2_1}{2\hbar(\omega-\omega_e)}~,
\end{equation}
which yields
\begin{equation}
\frac{K_{0,1}^2}{(\omega-\omega_e)}
\frac{NA}{2k_0}
=
-K_{0,0}R~.
\end{equation}
The quantum noise term can be written as
\begin{equation}
\begin{aligned}
K_{0,0}\Delta B^{\dag} (1,0,t)
&=
cK_{0,0}\left(\frac{LA\hbar}{2\pi^3\omega_e\mu_1^2}\right)^{\frac{1}{2}}\int_{-\infty}^{\infty} dk~ i\sin(kl)e^{i(\omega_k-\omega_e)t}
r_{k}^{\dag}(0)~.
\end{aligned}
\end{equation}
Using the relation
\begin{equation}
\begin{aligned}
cK_{0,0}\left(\frac{LA\hbar}{2\pi^3\omega_e\mu^2}\right)^{\frac{1}{2}}
=
\sqrt{\frac{2cK_{0,0}}{\pi}}
\left(\frac{L}{4\pi}\right)^{\frac{1}{2}}
\frac{\mu_0}{\mu_1},
\end{aligned}
\end{equation}
we obtain the delay differential operator equation for the polarization
\begin{equation}
\begin{aligned}
\frac{d}{dt}P^{\dag}(t)
&=
-K_{0,0} P^{\dag}(t)-K_{0,0}Re^{-i\omega_e \tau}P^{\dag}(t-\tau)\Theta(t-\tau)
\left(
G(t)-E(t)
\right)\\
&\quad+
R_{\mathcal{N}}\int_{-\infty}^{\infty} dk~ i
\sqrt{\frac{2cK_{0,0}L}{\pi}}
\sin(kl)e^{i(\omega_k-\omega_e)t}
r_{k}^{\dag}(0)\left(
G(t)-E(t)
\right)~,
\end{aligned}
\label{eq:FINAL}
\end{equation}
where $R_{\mathcal{N}}:=\mu_0/(\mu_1\sqrt{4\pi})$.\\
This equation is similar to the Eq.~\eqref{eq:DelayEff} but differs in 
the appearance of the reflection coefficients $R$ and $R_{\mathcal{N}}$. 
As for perfect mirrors $R=-1$ holds, equation \eqref{eq:DelayEff} and
\eqref{eq:FINAL} are identical for perfect mirrors.
Further, we emphasize that Eq.~\eqref{eq:FINAL} only holds for $|R|\ll 1$. 
However, the results calculated using the model described 
in Sec.~\ref{sec:HamEff} can be generalized to higher reflectivities,
including counter-rotating contributions in the Hamiltonian. 
The following graph Fig.~\ref{Fig:2} depicts qualitatively the behavior of the expectation value $\langle E\rangle$ for different $R$ on $[0,2\tau)$.
Our derivation allows to choose a reflectivity, and we see the influence
of the mirror properties in the degree of feedback that is measurable
on the emitter's dynamics.
For $ R=-1$, the dielectric is a perfect mirror, and the usual 
results from \cite{kabuss2016unraveling,dorner2002laser} are rederived,
however, now within a microscopic model.
This is the main result of the paper.
Since now the mirror dynamics are explicitly included, we can also start to investigate
regimes, where the mirror is not a passive optical element anymore but
is for example driven.
This allows to include gain into the system.
Just by hand, we choose a $ R<-1 $ to visualize, how the feedback will
be changed due to gain.
We stress, that the gain dynamics is just indicated.
\section{Conclusion} \label{sec:Concl}
The presented microscopic approach to quantum feedback yields a delay 
differential operator equation for the polarization $P^{\dag}$ Eq.~
\eqref{eq:GLEICHUNG} including dielectric characteristics. 
This expands the analysis of quantum coherent time-delayed feedback to a 
wider class of mirrors, e.g., metallic, dielectric, active, passive etc. 
Using an equation of motion approach we provide the possibility to 
describe an excited reflecting dielectric under stimulated emission, i.e., a
description of quantum optical gain in mirror-emitter setups. 
In addition, the presented analysis gives access to a wider class of active quantum coherent feedback control where the intrinsic mirror properties are externally steered. 
This could have significance both in tailored control of external quantum emitters and multi-photon selective reflection properties. 
Furthermore, we have shown that the microscopic approach justifies the $\sin(kl)$ like coupling used in the effective Hamiltonian Eq.~\eqref{eq:HamEff}.
In conclusion, the microscopic approach is very promising for further investigations as it takes properties of the dielectric into account and does not contradict the qualitatively motivated model. 
\bibliographystyle{tfp}
\bibliography{lib}
\appendix
\section{Refraction Index} \label{App: A}
The microscopic approach to quantum feedback presented in this article makes extensive use of the rotating wave approximation.
Subsequently, we deduce the Fresnel coefficients in a framework where the approximation is valid.
We start with the general equality relating the refraction index $n$ and the susceptibility $\chi$ 
\begin{equation}
n^2-1=\chi~.
\end{equation}
Characterizing the 
susceptibility  
for a two-level system by using the dipole density, expressed for localized atomic levels yields
\begin{equation}
P(r,t)=\sqrt{\pi}\mu_1\langle \sigma_{1,2}\rangle N+\sqrt{\pi}\mu_1\langle \sigma_{2,1}\rangle N~.
\end{equation}
The density matrix element is described by
\begin{equation}
\frac{d}{dt}\langle \sigma_{l,m}\rangle
=
i\omega\langle \sigma_{l,m}\rangle-K_{1,1}-i\sum_{n\in\{1,2\}}\left(
\Omega^*_{l,n}\langle \sigma_{n,m}\rangle-\Omega_{m,n}\langle \sigma_{l,n}\rangle
\right)~,
\end{equation}
where $\Omega_{m,l}=\sqrt{\pi}\mu_1E(R,\omega_e)/\hbar$ is the corresponding matrix element of the 
Rabi frequency.
Since in linear optics the polarization dynamics is only linear in the field, we obtain
\begin{equation}
\frac{d}{dt}\langle \sigma_{l,m}\rangle
=
i\omega\langle \sigma_{l,m}\rangle-K_{1,1}-i\left(
\Omega^*_{l,m}\langle \sigma_{m,m}\rangle-\Omega_{m,l}\langle \sigma_{l,l}\rangle
\right)~.
\label{eq:sigmaFourier}
\end{equation}
Transforming to Fourier space, equation \eqref{eq:sigmaFourier} can be solved by
\begin{equation}
\langle \sigma_{l,m}\rangle
=
\frac{\sqrt{\pi}\mu_1E(R,\omega_e)}{\hbar}\frac{-i(\langle\sigma_{m,m}\rangle-\langle\sigma_{l,l}\rangle)}{-i(\omega+\omega_e)+K_{1,1}}~.
\end{equation}
This yields the dipole density
\begin{equation}
P(r,\omega_e)
=
\sum_{l,m\in\{1,2\}}
\sqrt{\pi}\mu_1\frac{\sqrt{\pi}\mu_1E(R,\omega_e)}{\hbar}\frac{-i(\langle\sigma_{m,m}\rangle-\langle\sigma_{l,l}\rangle)}{-i(\omega+\omega_e)+K_{1,1}}N~.
\end{equation}
We emphasize that for the sake of consistency we set $\varepsilon_0=1$.
As $P=\chi E$ we deduce
\begin{equation}
\begin{aligned}
\chi(\omega_e)
&=
N
\sum_{l,m\in\{1,2\}}
\frac{\pi\mu^2_1}{\hbar}\frac{(\langle\sigma_{m,m}\rangle-\langle\sigma_{l,l}\rangle)}{(\omega+\omega_e)-iK_{1,1}}\\
&=
\frac{N\pi\mu^2_1(\langle\sigma_{1,1}\rangle-\langle\sigma_{2,2}\rangle)}{\hbar}
\left(
\frac{(\omega+\omega_e)+iK_{1,1}}{(\omega+\omega_e)^2+K^2_{1,1}}
-
\frac{(\omega_e-\omega)-iK_{1,1}}{(\omega_e-\omega)^2+K^2_{1,1}}
\right)~.
\end{aligned}
\end{equation}
We neglect the term containing $(\omega+\omega_e)$ as it can not be resonant.
This yields
\begin{equation}
\chi(\omega_e)
=
\frac{N\pi\mu^2_1(\langle\sigma_{1,1}\rangle-\langle\sigma_{2,2}\rangle)((\omega-\omega_e)-iK_{1,1})}{\hbar((\omega-\omega_e)^2+K^2_{1,1})}~.
\label{eq:Chi}
\end{equation}
Considering an off-resonant setting, we can assume that the part of $\chi$ describing the absorption is close to zero.
This part is identified with the imaginary part of Eq.~\eqref{eq:Chi}.
Assuming $K_{1,1}\ll |\omega-\omega_e|$ we obtain
\begin{equation}
n^2(\omega_e)-1\approx\text{Re}(\chi(\omega_e))
\approx
\frac{N\pi\mu^2_1}{\hbar(\omega-\omega_e)}~.
\end{equation}
\section{Dual-Pairing of Delta-Distribution over Domians not Containing Zero} \label{App: delta} 
In this section we show how to compute the dual-pairing of the delta-distribution with functions over domains not containing zero in physical systems. 
The idea is to expand the function in the dual pairing with a Heaviside step function.
This yields the problem that the step function can be defined in zero with any number $\alpha\in[0,1]$ multiplied by the function in the dual pairing evaluated in zero.
We show that $\alpha$ has to be $\alpha=1/2$ for quantum optical systems, otherwise we would contradict well known commutator relations.\\
We start with the qualitatively justified Hamiltonian Eq.~\eqref{eq:HamEff} and deduce for $t\in[0,\tau)$ the operator
\begin{equation}
\begin{aligned}
P^{\dag}(t)
&=e^{i\omega_e t}e^{-\kappa t}P^{\dag}(0)+ig_0\int_{0}^tdt'e^{ \kappa(t'-t)}\int_{\mathbb{R}}dk~
e^{-i((\omega_e-\omega_k)t'-\omega_e t)}r^{\dag}_{k}(0)\Delta(t')~,
\end{aligned}
\end{equation}
where $\kappa=g^2_0\pi\alpha/c$ and $\Delta(t')=G(t')-E(t')$. 
We find a similar solution for $P$. 
Assuming a system with weak decay yields $\Delta(t')\approx -1$.
We then find
\begin{equation}
\,[P(t), P^{\dag}(t)]_+
=
e^{-2\kappa t}+\frac{g_0^2\pi }{2c\kappa}
\left(1-e^{ -2\kappa t}\right)~.
\end{equation}
As $[P(t), P^{\dag}(t)]_+=G+E=1$ we know that
\begin{equation}
\frac{g_0^2\pi }{2c\kappa}\overset{!}=1~.
\end{equation}
Inserting the definition of $\kappa$ yields
\begin{equation}
\frac{g_0^2\pi }{2c\kappa}
=\frac{cg_0^2\pi }{2cg^2_0\pi\alpha}
=\frac{  1}{2 \alpha}~.
\end{equation}
Hence, the only value of $\alpha\in[0,1]$ for which $[P(t), P^{\dag}(t)]_+=G+E=1$ is $\alpha=1/2$.

\end{document}